\documentclass[aps,prb,twocolumn,groupedaddress,showpacs]{revtex4}
\usepackage{graphicx}
\usepackage{dcolumn}
\usepackage{bm}
\usepackage{color}

\newcommand{\figwidth}{0.95\columnwidth}

\newcommand{\figwidthhalf}{0.43\columnwidth}

\begin{document}

\title{Point Mutations Effects on Charge Transport
Properties of the Tumor-Suppressor Gene $p53$}

\author{Chi-Tin Shih$^\dag$, Stephan Roche$^\ddag$, Rudolf A. R\"{o}mer$^*$}
\affiliation{$^\dag$Department of Physics, Tunghai University,
40704 Taichung, Taiwan\\ $^\ddag$CEA/DSM/DRFMC/SPSMS, 17 avenue
des Martyrs, 38054 Grenoble, France\\$^*$Department of Physics and
Centre for Scientific Computing, University of Warwick, Gibbet
Hill Road, Coventry, CV4 7AL, UK}



\date{$Revision: 1.11 $, compiled \today}

\begin{abstract}   We report on a theoretical study of point mutations effects on charge
  transfer properties in the DNA sequence of the tumor-suppressor p53
gene.
  On the basis of effective single-strand or double-strand tight-binding models which
  simulate hole propagation along the DNA, a statistical
analysis of
  charge transmission modulations associated with all possible point
  mutations is performed. We find that in contrast to non-cancerous mutations,
  mutation hotspots tend to result in significantly weaker
  {\em changes of transmission properties}. This suggests that charge
transport could play a significant role for DNA-repairing
deficiency yielding carcinogenesis.
\end{abstract}

\pacs{87.15.Aa, 87.14.Gg, 87.19.Xx}

\maketitle

\begin{figure}{\includegraphics[width=\figwidth]{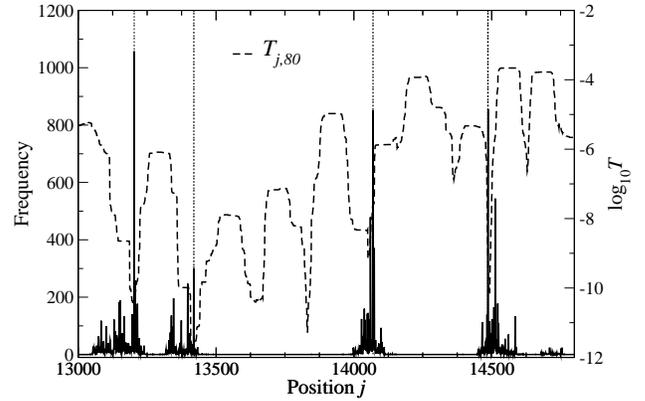}}
\caption{Mutation frequency of each site (thin lines) and
    averaged transmission coefficient $\bar{T}_{j,80}$ (dashed line).
Vertical dotted lines denote known regions of frequent
mutations (hotspots).} \label{fig:hot}
\end{figure}

The charge transfer properties and long range oxidation mechanisms
in DNA molecules are believed to play a critical role in the
living organisms.\cite{EndCS04,Cha07} For instance, it is believed
that base excision repair (BER) enzymes locate the DNA base
lesions or mismatches by probing the DNA-mediated charge transport
(CT). \cite{Raj00,YavBSB05} The $p53$ DNA is said to be the
``guardian of the genome'' since it encodes the $TP53$ protein
that suppresses the tumor development by activating the DNA repair
mechanisms or the cell apoptosis process if the damage of DNA is
irreparable. More than $50\%$ of human cancers are related to the
mutations of the $p53$ gene which jeopardize the efficient
functioning of $TP53$. \cite{She04} Most of the cancerous
mutations are point mutations
--- a base pair substituted by another --- with distributions
along the DNA sequence that are highly non-uniform.\cite{PetMKI07} The positions where the mutations occur most
frequently are call the ``hotspots'' of mutations. Each point
mutation can be characterized by two parameters $k$ and $s$,
respectively representing the position of the mutation on the
sequence and the nucleotide substituting the original one. From
the IARC database,\cite{PetMKI07} one finds that most hotspots of
$p53$ are located in the exons 5, 6, 7, and 8 in the interval from
the $13055$th to the $14588$th nucleotide. The distribution of the
point mutations in this range is reported in Fig.\ \ref{fig:hot}.
In this Letter, by using single and double strands tight-binding
models with parameters fitted from ab initio
calculations\cite{Cha07,CunCPD02}, the charge transmission changes
owing to cancerous and non-cancerous point mutations are
statistically studied for the p53 gene. We find that anomalously
small changes of charge transfer efficiency modulations coincide
with cancerous mutations. In contrast, non-cancerous mutations
result, on average, in much larger changes of the CT properties.
From this analysis, we suggest a new scenario how cancerous
mutations could shortcut the DNA damage/repair processes and hence
yield carcinogenesis.

\begin{figure}
  (a)\includegraphics[width=\figwidthhalf]{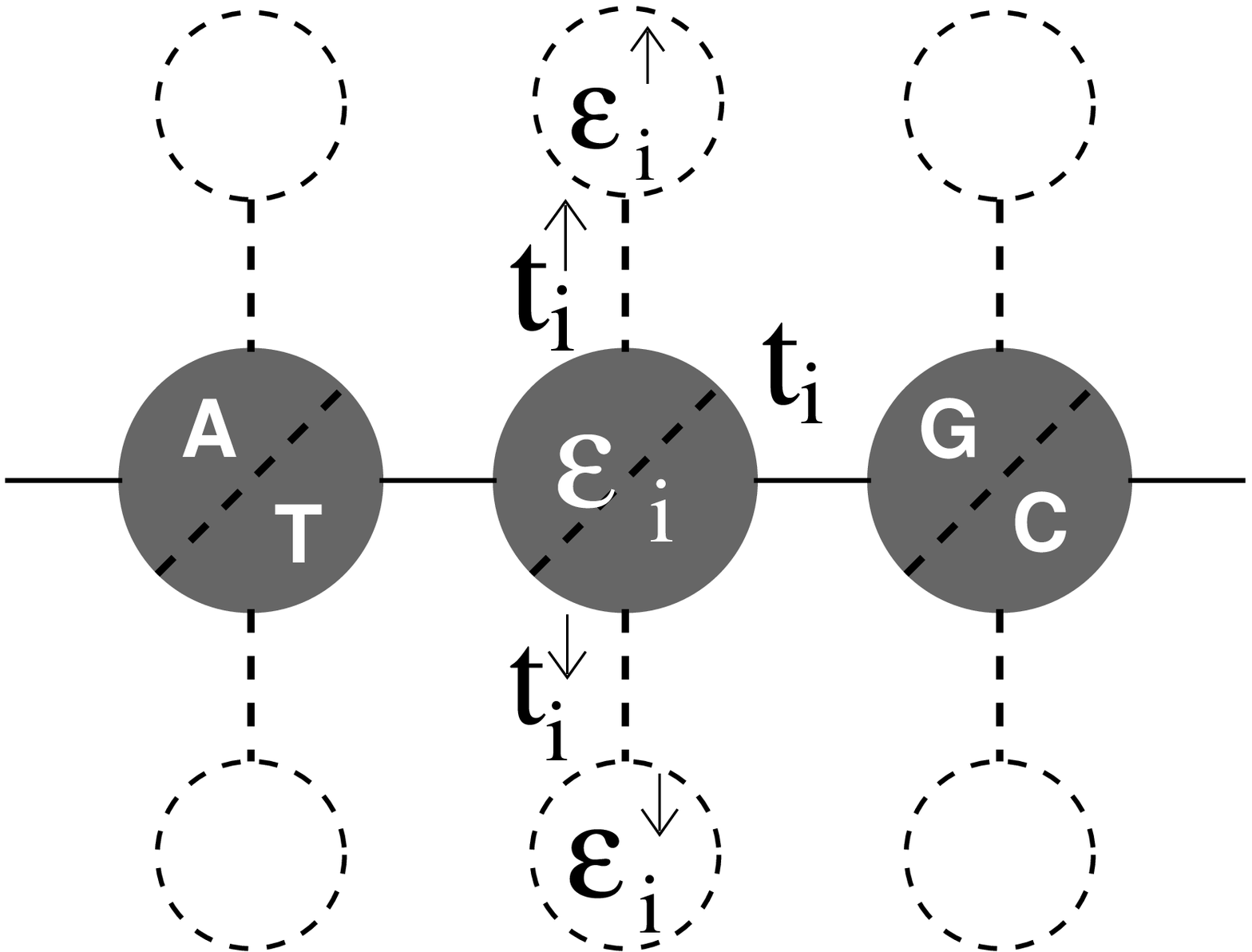}
  (b)\includegraphics[width=\figwidthhalf]{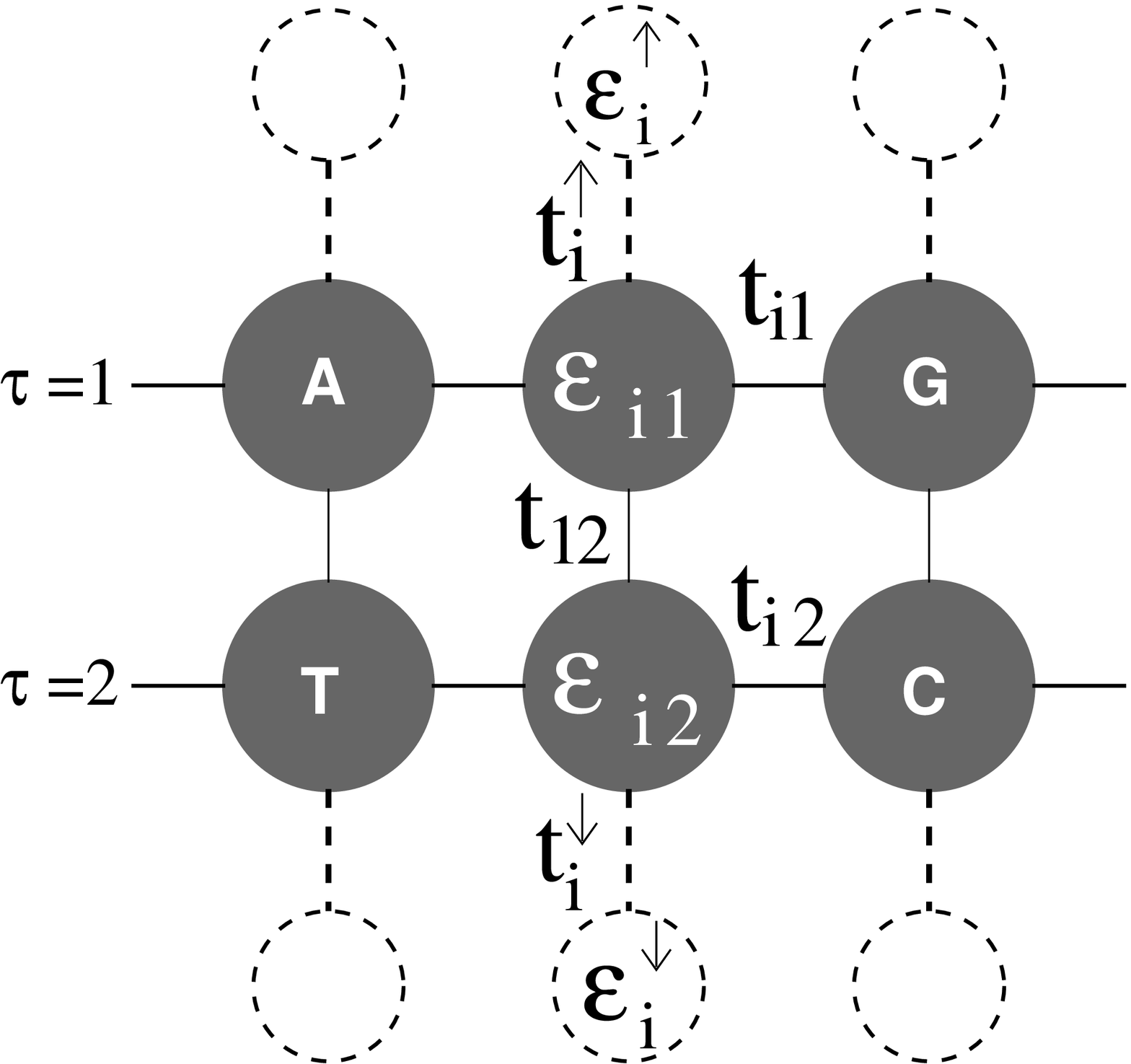}
  \caption{\label{fig:models} Schematic models for hole
    transport in DNA. The nucleobases are given as (grey) circles.
    Electronic pathways are shown as lines, and dashed lines and circles
    denote the sugar-phosphate backbone. Graph (a) shows effective
    models 1L and FB (with dashed backbone) for transport along a
    single channel, whereas graph (b) depicts possible two-channel
    transport models 2L and LM (with dashed backbone). }
\end{figure}

A simple but physically reasonable description of coherent hole
transport in single strand DNA is given by an effective tight-binding
Hamiltonian\cite{BerBR02}
\begin{equation}
  H=\sum_n \epsilon_nc_n^{\dagger}c_n - \sum_n t_{n,n+1}(c_n^\dagger
  c_{n+1} + h.c.)
\label{eq:model}
\end{equation}
where each lattice point represents a nucleotide base (A,T,C,G) of
the chain for $n=1, \ldots, N$. This one-leg (1L) model is shown
schematically in Fig.\ \ref{fig:models}(a).
In this tight-binding
formalism, $c_n^\dagger$ ($c_n$) is the creation (destruction)
operator of a hole at the $n$th site. The $t_{n,n+1}$ are the
hopping integrals along the DNA. $\epsilon_n$ is related to the
ionization potential at the $n$th site.
The electronic energetics of a DNA chain should take into account
three different contributions coming from the nucleobases system,
the backbone system and the environment.\cite{Cha07} We
emphasize that in many of the models to be used here, simplified
assumptions about these energy scales have to be employed. Mostly,
however, the ionization energies\index{ionization energies}
$\epsilon_{\rm G}=7.75 e$V, $\epsilon_{\rm C}=8.87 e$V,
$\epsilon_{\rm A}=8.24 e$V and $\epsilon_{\rm T}=9.14 e$V,
\cite{VoiJBR01} are taken as
suitable approximations for the onsite energetics at each base as
well as $7.75 e$V for the electrodes.\cite{BerBR02,Roc03,VoiJBR01,Shi06a,DiaSSD07}
Furthermore, in the 1L model $t_{n,n+1}$ is assumed to be
nucleotide-independent with $t_{n,n+1}=0.4e$V following prior
modelling in agreement with ab initio calculations.\cite{BerBR02}

A straightforward generalization of model (\ref{eq:model})
includes a two-leg ladder model (2L) as shown in Fig.\
\ref{fig:models} (b). The hopping between like base pairs (AT/AT,
GC/GC, etc.) is chosen as $0.35 e$V, between unlike base pairs it
is $0.17 e$V; the interchain hopping $t_{\perp}= 0.1 e$V. Other
models \cite{Cha07} include the presence of sites which represent
the sugar-phosphate backbone of DNA but along which no electron
transport is allowed (cp.\ Fig.\ \ref{fig:models}). In the
following we call the one-channel variety a {\em fishbone} (FB)
and the two-channel version {\em ladder model} (LM). The
additional hopping onto the backbone is $0.7 e$V and the backbone
onsite energy is taken to be $8.5 e$V, roughly equal to the mean
of all onsite energies for the base pairs.

The most convenient method for studying the transport
properties of these $4$ quasi-one-dimensional tight-binding models
is the transfer-matrix method,\cite{PicS81a} which allows
us to determine the transmission coefficient $T(E)$ of hole states
in systems with varying cross section $M$ and length $L \gg M$.
Briefly, we can solve for the eigenstates $|\Psi\rangle = \sum_n
\psi_n| n\rangle$ of the Hamiltonian, where $|n\rangle$ represents
the state that the hole is located in the $n$th site, as
$\left(\psi_{L}, \psi_{L-1}\right)^T = \tau_L\cdot \left( \psi_1,
\psi_0 \right)^T$ where $\tau_L (E)$ is the global transfer
matrix.\cite{PicS81a} $E$ is the energy of the injected carrier.
The transmission $T(E)$ is given in terms of $\tau_L(E)$ by a
simple analytic formula \cite{Mac99} for the 1L and FB models and
can be computed from the localization lengths for the 2L and LM
models.\cite{PicS81a}


Let us define $S=(s_1,s_2,\cdots,s_{20303})$ as the sequence of
the $p53$ gene (NCBI access number $X54156$, $20303$ base
pairs),\cite{Fut91} whereas $S_{j,L}$ is a segment of $S$ with
length $L$ starting at the $j$th base pair, i.e.
$S_{j,L}(n)=S(j-1+n)$ with $n=1,2,\ldots,L$. Next, we denote by
$T_{j,L}(E)$ the transmission coefficient corresponding to
$S_{j,L}$. We then characterize the energy-averaged CT for the
$j$th site with segment length $L$ as the value $\bar{T}_{j,L}$
obtained by integrating $T_{j,L}(E)$ for all incident energies and
all possible $L$ subsequences of all $p53$ segments of length $L$
containing the $j$th site such that
\begin{equation}
  \bar{T}_{j,L}=\frac{1}{L}\sum_{n=j-L+1}^{j}
  \frac{1}{E_1-E_0}\int^{E_1}_{E_0} T_{n,L}(E) dE
\label{eq:ave_te}.
\end{equation}
where $n$ is further restricted to $1\leq n \leq 20304-L$ close to
the boundaries; $E_0$ and $E_1$ denote a suitable energy window
which we shall normally choose to equal the extrema of the energy
spectrum for each model.
In Fig.\ \ref{fig:hot} we show $\bar{T}_{j,80}$ for model 1L and
base pair range $13000<j<14800$ where the most cancerous mutations
occur. The positions of four groups of hotspots, i.e.\ peaks of
the mutation frequency, corresponding to the four exons
($5$-$8$th) coincide with local minima of $\bar{T}_{j,80}$.

If the $k$th base on the $p53$ sequence is mutated from $s_k$ to
$s$ and $j\le k\le j+L-1$, we will denote the mutated segment
containing this mutation as $S^{k,s}_{j,L}$ such that
$S^{k,s}_{j,L}(k-j+1)=s$ and $S^{k,s}_{j,L}(i)=S_{j,L}(i)$ for all
$i\neq k-j+1$. The corresponding transmission coefficients of the
original and mutated sequence are denoted as $T_{j,L}(E)$ and
$T^{k,s}_{j,L}(E)$, respectively. Similarly, we define the
energy-averaged squared differences in transmission coefficient
between original and mutated sequence as
\begin{equation}
  \bar{\Delta}^{k,s}_{j,L}=\frac{1}{E_1-E_0}\int^{E_1}_{E_0}
  \left|T_{j,L}(E)-T^{k,s}_{j,L}(E)\right|^2 dE
\label{eq:delta}.
\end{equation}

\begin{figure}
  \includegraphics[width=\figwidth]{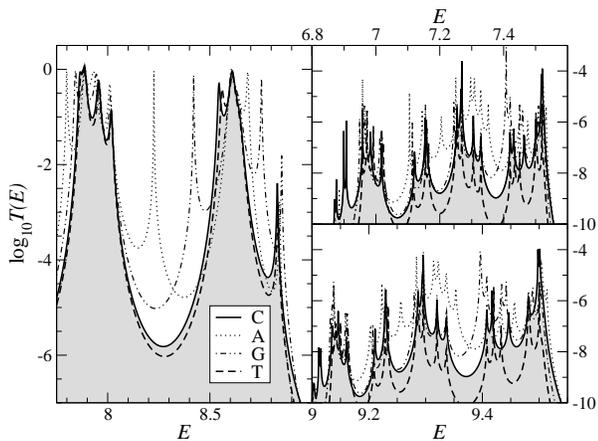}
  \caption{
    Energy-dependence of logarithmic transmission coefficients
    $T^{14585,s}_{14575,20}(E)$ of the original sequence ($C$
shaded solid line)
    and mutated ($A$ dotted, $G$ dotted-dashed, $T$ dashed)
sequences with length
    $L=20$ (from $14575$th to $14594$th nucleotide) of $p53$. The left
    panel shows results for model 1L, the right two panels denote the
    two transport windows for the fishbone model.\cite{KloRT05}}
  \label{fig:mut}
\end{figure}

The $14585$th base pair 
of the $p53$
sequence is a particularly active hotspot with $133$ entries in
the IARC database.\cite{PetMKI07} It exhibits mutations from $C$
to $T$ and causing various types of cancer. However, the mutations
$C \to G$ and $C\to A$ at the same position are not cancerous.
The effects of the cancerous $C\to T$ and the non-cancerous $C\to
A$, $C\to G$ mutations on the CT properties are shown in
Fig.~\ref{fig:mut}
.
The transmission coefficients $T_{14575,20}(E)$ and
$T^{14585,s}_{14575,20}(E)$ with $s=T$, $A$ and $G$ are given in
Fig.~\ref{fig:mut}. We find that for most energies the mutation
$C\to T$ results in the weakest change in $T(E)$.

To evaluate the change of CT for {\it all} mutations in p53
quantitatively, $\bar{\Delta}^{14585,s}_{14575,20}$ are computed
for all the four models. The results are shown in Table
\ref{tab:Delta}.
\begin{table}
  \centering
\caption{Renormalized values of the energy-averaged changes
$\bar{\Delta}^{14585,s}_{14575,20}$ in transmission properties for
the $4$ tight-binding models. All data are shown with at most 3
significant figures. Common multiplication factors for each group
of data for given $L$ and mutations with $C\to A$, $G$ and $T$ are
suppressed. Bold entries denote minima for the CT change.}
  \begin{tabular*}{\hsize}{@{\extracolsep{\fill}}cccccc}
    \hline\hline
    $s$      & $L$& 1L    & FB    & 2L   & LM \\ \hline
    $C\to A$ & 20 & 23.1 & 8.46  & 2.24 & {\bf 0.43} \\ \hline
    $C\to G$ & 20 & 37.6  & {\bf 0.73}  & 0.83 & 0.57 \\ \hline
    $C\to T$ & 20 & {\bf 5.63}  & 1.08  & {\bf 0.34} & 0.66 \\ \hline
\hline
    $C\to A$ & 30 & 15.7  & 54.8  & 96.2 & 1.76 \\ \hline
    $C\to G$ & 30 & 21.4  & 0.55  & 2.75 & 0.40 \\ \hline
    $C\to T$ & 30 & {\bf 9.14}  & {\bf 0.0006}& {\bf 0.39} & {\bf 0.15}
\\ \hline \hline
    $C\to A$ & 40 & 1.16  & 30.7  & 31.6 & 17.7 \\ \hline
    $C\to G$ & 40 & 2.21  & 0.72  & 0.41 & 0.16 \\ \hline
    $C\to T$ & 40 & {\bf 0.40}  & {\bf 0.009} & {\bf 0.26} & {\bf 0.04}
\\ \hline \hline
\end{tabular*}
\label{tab:Delta}
\end{table}
We see that the cancerous mutation $C\to T$ shows the smallest
relative change in CT for nearly all models. The only differences
occur for small $L=20$ in models FB and LM but vanish quickly for
larger $L$. Hence for a damage-repair process which uses a
CT-based criterion as a detection mechanism, this mutation will be
the hardest to identify. These results seem to suggest a scenario
in which certain mutations might avoid the CT-driven DNA
damage-repair mechanism and survive to develop cancerous tumors.
We have checked that this trend is independent of the specific
model and hotspot chosen by analysing also the hotspots $13117$,
$13203$, $13334$, $13419$, $14060$, $14069$, $14070$, $14074$,
$14076$, $14486$, $14487$, $14501$, $14513$, and $17602$ of the
IARC TP53 data base \cite{PetMKI07} for DNA segment lengths $L=10,
20, \ldots 160$. We find that the number of cases in which a
cancerous mutation corresponds to a segment of low transmission
change is within $5\%$--$15\%$ the same for models L1, FB, L2 and
LM, with results for  L1 and FB very similar to
each other. The models L2 and LM are within $15\%$ of each other
and have only a slightly smaller occurrence of these cases of low
transmission change and high cancerousness than L1 and FB. Thus in
the following, we shall restrict our analysis to the simple case
of the strictly 1D model 1L given by (\ref{eq:model}).

Experimentally, the BER enzymes can locate the damaged sites at a
distance of $19$ base pairs on the DNA strand by probing the CT of
the segment bound by the enzymes.\cite{YavBSB05} If a
mutation changes the CT only slightly, the enzymes might thus not
be able to find it and the repair mechanism will not be activated.
On the other hand, the fact that mutations $C\to G$ and $C\to A$
are not found in cancer cells does not mean that these mutations
do not occur. Rather, the changes in CT induced by them are more
significant which could allow an easier detection by CT-probing
enzymes. Accordingly these two types of mutations will be repaired
and cancer will not develop.

\begin{figure}
  \centering
\includegraphics[width=\figwidth]{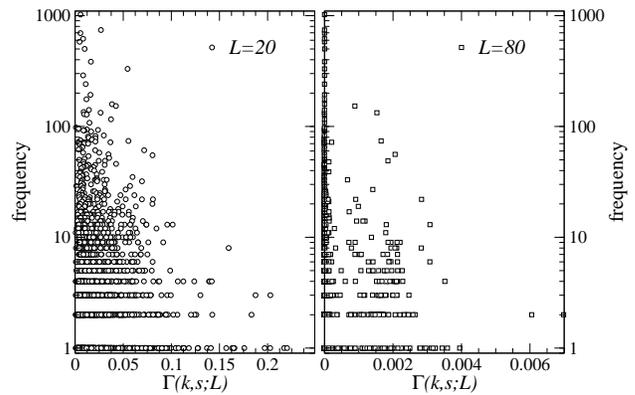}
  \caption{Scatter plots of $\Gamma(k,s;w)$ versus occurrence frequency
of all cancerous mutations $s$ corresponding to the hotspots
$\{k\}$ for (a) $L=20$ and (b) $80$.} \label{fig:scatt}
\end{figure}

In order to challenge such a scenario, the change of CT for all
$20303\times 3=60909$ types of possible point mutations are
examined. The average effect of a mutation $(k,s)$ of a
subsequence with length $L$ on the CT  of $p53$ is defined as
\begin{equation}
\Gamma(k,s;L)=\frac{1}{L}\sum_{j=k-L+1}^{k}\bar{\Delta}^{k,s}_{j,L}
\label{eq:gamma}
\end{equation}
where $j$ also satisfies $1\le j \le 20304-L$ close to boundaries.
Fig.\ \ref{fig:scatt} shows the scatter plots of $\Gamma(k,s;L)$
versus frequency of all cancerous mutations for (a) $L=20$ and (b)
$80$. The sharp peaks at small $\Gamma$ agree with the scenario
that the most cancerous mutations --- namely those with high
frequency --- change the CT only slightly and thus have smaller
$\Gamma$.

Let us now compare the CT change (i) for the set ${\cal M}$ of all
$60909$ possible point mutations of $p53$ (ii) for the set ${\cal
M}_{\rm c}$ of the $1953$ cancerous point mutations in the IARC
database \cite{PetMKI07} and (iii) for the set ${\cal M}_{{\rm
c},10}$ of the $366$ mutations which are found more than $10$
times in the cancer tissues.
For given $L$, we sort the CT results for ${\cal M}$ according to
the computed magnitude of $\Gamma(k,s;L)$ and determine the rank
$r(k,s;L)\in [1, 60909]$ of the CT change for each mutation
$(k,s)$. A smaller rank means less CT change for the mutation.
$\gamma(k,s;L)=100\%\times r(k,s;L)/60909$ is then the relative
rank in percentage.

\begin{figure}
  \centering
  \includegraphics[width=\figwidth]{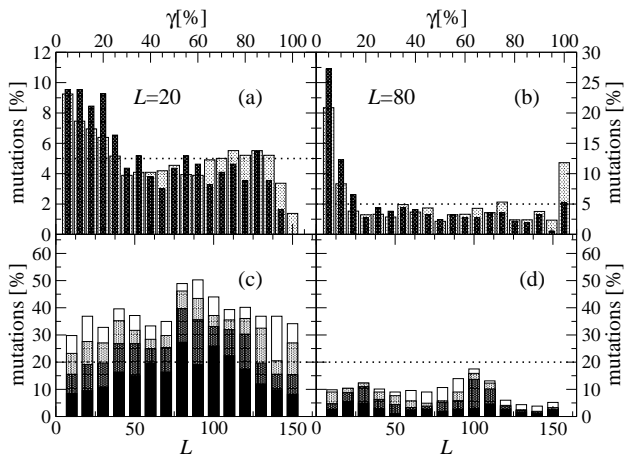}
  \caption{Histogram of the distribution of $\gamma(k,s;L)$ 
  in ${\cal M}_{\rm c}$
(light wide bars) and
    ${\cal M}_{{\rm c},10}$ (dark thin bars) which changes the $k$th
nucleotide to $s$ for (a)
    $L=20$ and (b) $80$. For ${\cal M}$, all values are equal to $5\%$
in the
    $20$ intervals as indicated by the horizontal dashed lines.
    (c) shows  the percentage of $\Gamma(k,s;L)$ values in ${\cal
M}_{{\rm c},10}$ for
    small CT change as a function of DNA lengths in the range $0$--$5\%$
(black), $5$--$10\%$ (dark grey), $10$--$15\%$ (light grey) and
$15$--$20\%$ (white). Similarly, (d) indicates large
    CT change for ${\cal M}_{{\rm c},10}$ in the ranges $80$--$85\%$
(black), $85$--$90\%$ (dark grey), $90$--$95\%$(light grey) and
$95$--$100\%$ (white). The horizontal dashed lines in (c) and (d)
indicates the distributions for ${\cal M}$.} \label{fig:ac_histo}
\end{figure}

%
The histograms of the distribution of $\gamma(k,s;L)$ are shown in
Fig.\ \ref{fig:ac_histo} (a) and (b) for the mutations $(k,s)$ of
${\cal M}_{\rm c}$ and ${\cal M}_{{\rm c},10}$. The vertical axis
is the percentage of mutations in ${\cal M}_{\rm c}$ (grey bars)
and ${\cal M}_{{\rm c},10}$ (black bars) whose $\gamma(k,s;L)$
belong to the corresponding bin range with each width set to
$5\%$.
For ${\cal M}$, the result is the dashed line at a value of $5\%$.
The distributions for ${\cal M}_{\rm c}$ and ${\cal M}_{{\rm
c},10}$ are clearly biased to smaller values of $\gamma$,
especially for the $L=80$ case. E.g.\ there are about $9\%$ for
$L=20$ and $27\%$ of mutations for $L=80$ in the ${\cal M}_{{\rm
c},10}$ set whose $\gamma(k,s;L)$ values are smaller than $5\%$.
This indicates that the cancerous mutations in ${\cal M}_{\rm c}$
and ${\cal M}_{{\rm c},10}$ result in smaller CT changes than
non-cancerous ones. The distribution bias is more apparent in
${\cal M}_{{\rm c},10}$ than that in ${\cal M}_{\rm c}$ in
agreement with the choice of mutations.

Let us also evaluate the dependence of the CT change on different
$L$. Figs.\ \ref{fig:ac_histo} (c) and (d) show the accumulated
percentage of mutations in ${\cal M}_{{\rm c},10}$ whose
$\gamma(k,s;L)$ values are smaller than $20\%$ and larger than
$80\%$, respectively.
We see that around $L=90$, more than $50\%$ of mutations in ${\cal
M}_{{\rm c},10}$ change the CT less  than $20\%$, and the number
of cancerous mutations with an $80\%$ or more change in CT is much
less than average for all $L$.



In summary, we
find that (i) the conductance of hotspots of cancerous
mutations is smaller than that of other sites, (ii) on average the
cancerous mutations of the gene yield smaller changes of the CT in
contrast with non-cancerous mutations, (iii) the tendency in (ii)
is stronger in the set of highly cancerous mutations with
occurence frequency $>10$.
These results suggest a possible scenario of how cancerous
mutations might circumvent the DNA damage-repair mechanism and
survive to yield carcinogenesis. However, our analysis is only
valid in a statistical sense and we do observe occasional
non-cancerous mutations with weak change of CT. For these, other
DNA repair processes should exist and we therefore do not intend
to claim that the DNA-damage repair solely uses a CT-based
criterion. Still, our results exhibit an intriguing and new
correlation between the electronic structure of DNA hotspots and
the DNA damage-repair process.

Further studies should investigate how robust  our conclusions are
with regards to electron-phonon coupling effects, electronic
correlations, or metal/DNA contact
interactions\cite{DiaSSD07,KloRT05,WanC06,GutMCP06,Mal07,SenGFB03,
ConR00,Con05,WeiWCY05,Sta03,Mac05}. Since
mesoscopic transport measurements of DNA sequences of several tens
of base pairs have been demonstrated\cite{PorBVD00}, our theoretical results could be challenged by investigating charge transfer in wild and mutated short synthesized sequences of the p53 gene.

This work was supported by the National Science Council in Taiwan (CTS, grant
 95-2112-M-029-003-) and the UK Leverhulme Trust (RAR, grant
 F/00 215/AH).  Part of the calculations were performed at the
National Center for High-Performance Computing in Taiwan.



\end{document}